\documentclass[10pt,conference]{IEEEtran}
\IEEEoverridecommandlockouts

\usepackage{cite}
\usepackage{amsmath,amssymb,amsfonts}
\usepackage{algorithmic}
\usepackage{graphicx}
\usepackage{textcomp}
\usepackage{xcolor}
\usepackage{xspace}
\usepackage{soul}
\usepackage{url}
\usepackage{multirow}
\usepackage{hyperref}
\usepackage{tcolorbox}

\def\BibTeX{{\rm B\kern-.05em{\sc i\kern-.025em b}\kern-.08em
    T\kern-.1667em\lower.7ex\hbox{E}\kern-.125emX}}

\newcommand{\rqone}{\textit{How prevalent is library abandonment in the Maven ecosystem?}}
\newcommand{\rqtwo}{\textit{What are the release activity patterns in abandoned versus active libraries?}}

\begin{document}

\title{Understanding Abandonment and Slowdown Dynamics in the Maven Ecosystem}

\author{
    \IEEEauthorblockN{Kazi Amit Hasan, Jerin Yasmin, Huizi Hao, Yuan Tian}
    \IEEEauthorblockA{
        School of Computing, Queen's University \\
        Kingston, ON, Canada \\
        \{kaziamit.hasan, jerin.yasmin, huizi.hao, y.tian\}@queensu.ca
    }
    \and
    \IEEEauthorblockN{Safwat Hassan}
    \IEEEauthorblockA{
        University of Toronto \\
        Toronto, ON, Canada \\
        safwat.hassan@utoronto.ca
    }
    \and
    \IEEEauthorblockN{Steven H. H. Ding}
    \IEEEauthorblockA{
        McGill University \\
        Montreal, Quebec, Canada \\
        steven.h.ding@mcgill.ca
    }
}

\maketitle

\begin{abstract}

The sustainability of libraries is critical for modern software development, yet many libraries face abandonment, posing significant risks to dependent projects. This study explores the prevalence and patterns of library abandonment in the Maven ecosystem. We investigate abandonment trends over the past decade, revealing that approximately one in four libraries fail to survive beyond their creation year. We also analyze the release activities of libraries, focusing on their lifespan and release speed, and analyze the evolution of these metrics within the lifespan of libraries. We find that while slow release speed and relatively long periods of inactivity are often precursors to abandonment, some abandoned libraries exhibit bursts of high frequent release activity late in their life cycle. Our findings contribute to a new understanding of library abandonment dynamics and offer insights for practitioners to identify and mitigate risks in software ecosystems.

\end{abstract}


\section{Introduction}\label{intro}
Open-source software (OSS) libraries play a crucial role in modern software development, as they are frequently adopted as dependencies in millions of open-source and commercial software systems. These dependencies enable developers to build applications efficiently by leveraging pre-existing solutions. However, the sustainability of these libraries is often uncertain, as they typically depend on volunteer contributors who may discontinue due to various reasons~\cite{lin2017developer,forsgren20212020,kaur2022exploring}. This reliance on volunteer contributors makes libraries vulnerable to turnover and, ultimately, \textit{abandonment}, which presents risks such as unresolved bugs and unpatched security vulnerabilities, which can severely impact dependent systems~\cite{zimmermann2019small,miller2023we}. 

Most existing studies on OSS project abandonment focus on analyzing the impact and risk associated with project abandonment~\cite{valiev2018ecosystem,avelino2019abandonment,coelho2020github,cogo2021deprecation,ait2022empirical,miller2023we,miller2025understanding}. To the best of our knowledge, the recent study by Miller et al.~\cite{miller2025understanding} is the only one that proposes strategies to mitigate the risk brought by abandonment. Their study highlights the importance of proactive notifications from maintainers before abandonment occurs. However, this approach relies heavily on maintainers actively signaling their intent to abandon a project, which is not always feasible or reliable. This highlights a research gap in identifying more indicators of package abandonment that can warn dependent projects about potential risks. Our study seeks to address this gap through a data-driven investigation of libraries' release activities. The focus on release activities is driven by the intuition that abandoned projects may exhibit a slowdown in their release speed prior to abandonment.
To achieve our goal, we conduct a case study on libraries in the Maven ecosystem. Despite its popularity, many Maven projects have been abandoned over time, raising concerns about the ecosystem’s long-term sustainability~\cite{shen2025understanding}. Our study addresses the following two research questions: 
\begin{description}
 \item[\textbf{RQ1:}] \rqone
 \item[\textbf{RQ2:}] \rqtwo
\end{description}

The first question examines the severity and evolution of abandonment in the Maven ecosystem, and the second investigates release patterns to differentiate abandoned libraries from active ones. By answering these questions, our study provides insights into the extent of abandonment and early warning signs, offering guidance for developers in dependency management and promoting ecosystem resilience.

\noindent \textbf{Replication package:} The dataset and code for reproducing our empirical study results are available at: \url{https://github.com/RISElabQueens/analyzing-abandonment-and-slowdowns}.

\section{Data Preparation and Terminology}\label{dataset}
\subsection{Dataset and Filtering Process}
Our study uses the Maven Central Neo4j dataset~\cite{msrdata,10.1145/3643991.3644879}, which contains a comprehensive graph of the entire Maven ecosystem. In this study, we used the dataset updated till September 4, 2024. To prepare the data for analysis, we used Neo4j to extract release information of all libraries in the ecosystem. We observed that not all libraries have release data. After excluding those libraries, we retained 635,003 libraries (96.5\% of the original libraries). For our study, we define an \textit{observation window} spanning 10 years, from September 4, 2014, to September 4, 2024. This window focuses on relatively newer libraries whose complete lifecycles can be observed within this period. We excluded older libraries created before the start of the observation window, reducing the dataset to 581,424 unique libraries. Additionally, we excluded too new libraries (created for less than one year), and those with only a single release, as they lack sufficient data for meaningful analysis. After this filtering, our final dataset for answering two RQs contains 403,048 unique libraries, accounting for 61.3\% of the original libraries.


\subsection{Definitions}
We define the key terminology used in this study below.

\noindent \textbf{Abandoned and Active Libraries}: Libraries with no releases in the last two years of our observation window are classified as \textit{abandoned}; otherwise, they are considered \textit{active}. We acknowledge the threats associated with this definition and discuss them in detail in Section~\ref{threats}.

\noindent \textbf{Abandonment Date}: The date of the latest release for an abandoned library.

\noindent \textbf{Lifespan of a Library}: The time span between a library's first release and its latest release.

\noindent \textbf{Lifespan of a Release}: The time span between a target release and its subsequent release.

\noindent \textbf{Release Speed of a Library}: The average rate of releases over a specific time interval, measured in releases per day, month, or year. This metric was introduced by Damien et al.~\cite{jaime2022preliminary}.

\section{Prevalence of Library Abandonment}\label{rq1}

\subsection{RQ1.1: What is the trend of abandonment over the years? }

\begin{table*}[h]
\centering
\footnotesize
\setlength{\tabcolsep}{3pt} 
\renewcommand{\arraystretch}{0.75} 
\caption{Statistics of library abandonment in the maven ecosystem over 10 years (2014–2023).}
\vspace{-0.2cm}
\label{tab:year_wise_summary}
\begin{tabular}{ccccccc}
\hline
\textbf{Year} &
  \textbf{\begin{tabular}[c]{@{}c@{}}Continuing libraries \\ from previous year\end{tabular}} &
  \textbf{\begin{tabular}[c]{@{}c@{}}Newly created libraries \\ in current year\end{tabular}} &
  \textbf{\begin{tabular}[c]{@{}c@{}}Total active libraries \\ in current year\end{tabular}} &
  \textbf{\begin{tabular}[c]{@{}c@{}}Total abandoned \\ libraries in current year\end{tabular}} &
  \textbf{\begin{tabular}[c]{@{}c@{}}Abandonment rate \\ per year (\%)\end{tabular}} &
  \textbf{\begin{tabular}[c]{@{}c@{}}Cumulative abandoned \\ libraries\end{tabular}}  \\ \hline
2014 & 0      & 26,158 & 26,158  & 1,783   & 6.8\% & 1,783  \\ 
2015 & 24,375  & 34,620 & 58,995  & 12,354  & 20.9\% & 14,137 \\ 
2016 & 46,641  & 34,680 & 81,321  & 19,604  & 24.1\% & 33,741  \\ 
2017 & 61,717  & 34,405 & 96,122  & 21,523  & 22.3\% & 55,264 \\ 
2018 & 74,599  & 36,440 & 111,039 & 27,110  & 24.4\% & 82,374 \\ 
2019 & 83,929  & 40,550 & 124,479 & 28,075  & 22.5\% & 110,449 \\ 
2020 & 96,404  & 43,149 & 139,553 & 30,925  & 22.1\% & 141,374 \\ 
2021 & 108,628 & 63,612 & 172,240 & 37,655  & 21.8\% & 179,029  \\ 
2022 & 134,585 & 52,235 & 186,820 & 27,739  & 14.8\% & 206,768\\ 
2023 & 159,081 & 37,199 & 196,280 & 0  & 0.0\% & 206,768 \\ \hline
\end{tabular}
\vspace{-0.2cm}
\end{table*}

\noindent \textbf{Approach:} Following the definitions outlined in Section~\ref{dataset}, we classify libraries as either abandoned or active. For each library, we determine its entry into the ecosystem based on its first release date, and for abandoned libraries, we identify the date of abandonment. Subsequently, we collect key statistics related to abandonment for each observation year.


\noindent \textbf{Results:} Table~\ref{tab:year_wise_summary} summarizes, for each year, the number of active libraries, the libraries abandoned that year, and the percentage of total libraries abandoned. \textbf{We observed that as the Maven ecosystem rapidly expanded, the number of abandoned libraries also increased, with high annual abandonment rates consistently ranging from 20.9\% to 24.4\% during 2015 to 2021.} Note that the abandonment rate for 2014 is significantly lower at 6.8\%, likely because the observation window begins in 2014, providing only one year for libraries created in that year to be abandoned. Similarly, the lower abandonment rate in 2022 may be influenced by its proximity to the end of the observation window, which limits the time available for libraries to be classified as abandoned.


\subsection{RQ1.2: What is the abandonment rate of libraries over their life cycle?}

\begin{table*}[h]
\centering
\footnotesize
\setlength{\tabcolsep}{2pt} 
\renewcommand{\arraystretch}{0.75}
\caption{Summary of abandoned libraries based on their creation year.}
\vspace{-0.2cm}
\label{tab:survival_stats}
\begin{tabular}{ccccccccc}
\hline
\textbf{Year} &
  \textbf{Total libraries} &
  \textbf{\begin{tabular}[c]{@{}c@{}}Abandoned \\ libraries\end{tabular}} &
  \textbf{\begin{tabular}[c]{@{}c@{}}Active \\ libraries\end{tabular}} &
  \multicolumn{5}{c}{\textbf{\begin{tabular}[c]{@{}c@{}}Abandonment rates (\%)\end{tabular}}} \\ \cline{5-9}
     &       &       &       & \textbf{within 1 year} & \textbf{within 2 year} & \textbf{within 3 year} & \textbf{within 4 year} & \textbf{\begin{tabular}[c]{@{}c@{}}within 5 and  remaining years\end{tabular}} \\ \hline
2014 & 26,158 & 19,122 & 7,036  & 6.8\%   & 25.6\%  & 41.1\%  & 49.7\%  & 73.1\%  \\
2015 & 34,620 & 27,256 & 7,364  & 21.4\%  & 41.9\%  & 53.5\%  & 62.1\%  & 78.7\%  \\
2016 & 34,680 & 27,706 & 6,974  & 24.3\%  & 44.3\%  & 58.2\%  & 65.7\%  & 79.8\%  \\
2017 & 34,405 & 26,038 & 8,367  & 24.2\%  & 47.1\%  & 57.8\%  & 66.7\%  & 75.6\%  \\
2018 & 36,440 & 26,016 & 10,424 & 26.3\%  & 48.1\%  & 59.4\%  & 67.1\%  & 71.3\%  \\
2019 & 40,550 & 25,197 & 15,353 & 26.5\%  & 46.5\%  & 56.1\%  & 62.1\%  & --      \\
2020 & 43,149 & 22,212 & 20,937 & 24.9\%  & 42.7\%  & 51.4\%  & --      & --      \\
2021 & 63,612 & 25,263 & 38,349 & 26.7\%  & 39.7\%  & --      & --      & --      \\
2022 & 52,235 & 7,958  & 44,277 & 15.2\%  & --      & --      & --      & --      \\
2023 & 37,199 & 0      & 37,199 & --      & --      & --      & --      & --      \\ \hline
\end{tabular}
\end{table*}


\noindent \textbf{Approach:} RQ1.1 demonstrates that abandonment is prevalent in the ecosystem. However, the abandonment ratio presented in Table~\ref{tab:year_wise_summary} does not tell when a library will likely be abandoned, which motivates RQ1.2. To answer this question, we performed a cohort-based, time-to-event analysis. We define a \textit{cohort} as a group of libraries created in the same observation year. This design enables us to examine how abandonment patterns evolve. For each cohort, we calculated the proportion of libraries created in that year that were abandoned within 1 to 4 years, or later, from their creation date.


\noindent \textbf{Results:} The cohort-wise abandonment analysis (Table~\ref{tab:survival_stats}) shows varying abandonment patterns by library creation year. Libraries created in 2014 had the lowest first-year abandonment rate (6.8\%), but this rose sharply to 49.7\% within three years. \textbf{In contrast, libraries created between 2015 and 2021 showed higher first-year abandonment rates (21.4\%–26.7\%), with roughly one in four failing within their first year.} Second-year abandonment was particularly pronounced, driving the two-year cumulative rate to nearly 50\%. Notably, libraries created in 2018 and 2019 experienced the highest overall abandonment rates.






\vspace{-0.1cm}
\begin{tcolorbox}[colframe=white, colback=blue!5, left=1mm, right=1mm, sharp corners]
\textbf{RQ1 Summary:} Library abandonment is common in the Maven ecosystem, with a consistently high annual rate exceeding 20\%. Between 2015 and 2021, approximately 25\% of newly introduced libraries were abandoned within their first year, increasing to 39.7\%–48.1\% by the end of their second year.
\end{tcolorbox}


\section{Release Activity Patterns in Abandoned and Active Libraries}\label{rq2}



\subsection{RQ2.1: How do the lifespan and release speed of abandoned libraries compare to those of active libraries?}

\begin{table*}[]
\centering
\footnotesize
\setlength{\tabcolsep}{2pt} 
\renewcommand{\arraystretch}{1.1} 
\caption{Summary of the lifespan of libraries and release speed categories for active and abandoned libraries.}
\vspace{-0.2cm}
\label{tab:comparison_lifespan_speed}
\resizebox{\textwidth}{!}{%
\begin{tabular}{ccccc|cccc} 
\hline
\multirow{2}{*}{\textbf{Speed category}} &
  \multicolumn{4}{c}{\textbf{Abandoned libraries}} &
  \multicolumn{4}{c}{\textbf{Active libraries}} \\ \cline{2-9} 
 &
  \textbf{\textless{}1 year} &
  \textbf{1-2 years} &
  \textbf{\textgreater{}2 years} &
  \textbf{Total} &
  \textbf{\textless{}1 year} &
  \textbf{1-2 years} &
  \textbf{\textgreater{}2 years} &
  \textbf{Total} \\ \hline
\textbf{\textless{}1} &
  42,039 (20.3\%) &
  28,718 (13.9\%) &
  39,468 (19.1\%) &
  110,225 (53.3\%) &
  16,438 (8.4\%) &
  26,920 (13.7\%) &
  75,999 (38.7\%) &
  119,357 (60.8\%) \\ 
\textbf{1-2} &
  42,061 (20.3\%) &
  5,401 (2.6\%) &
  5,153 (2.5\%) &
  52,615 (25.5\%) &
  11,930 (6.1\%) &
  8,301 (4.2\%) &
  22,179 (11.3\%) &
  42,410 (21.6\%) \\ 
\textbf{\textgreater{}2} &
  37,907 (18.3\%) &
  3,312 (1.6\%) &
  2,709 (1.3\%) &
  43,928 (21.3\%) &
  10,771 (5.5\%) &
  7,394 (3.7\%) &
  16,348 (8.3\%) &
  34,513 (17.5\%) \\ \hline
\textbf{Total} &
  122,007 (59.1\%) &
  37,431 (18.1\%) &
  47,330 (22.8\%) &
  206,768 (100.0\%) &
  39,139 (19.9\%) &
  42,615 (21.7\%) &
  114,526 (58.3\%) &
  196,280 (100.0\%) \\ \hline
\end{tabular}%
}
\end{table*}

\noindent \textbf{Approach:} Release speed is a critical characteristic of libraries in a software ecosystem~\cite{jaime2022preliminary}. Thus, we first analyze the release activities of abandoned libraries by calculating this metric within the whole library lifespan. Recognizing that release speed may vary depending on a library's lifespan, we categorize libraries based on their lifespan and classify their release speeds into distinct groups. Specifically, we use 1 and 2 years as thresholds to define short-lived, moderate-lived, and long-lived libraries, and 1 and 2 releases per month to distinguish between fast, moderate, and slow release speeds. These thresholds are determined based on observations of the overall distribution of lifespan and release speed and their interpretability.


\noindent \textbf{Results:} Table~\ref{tab:comparison_lifespan_speed} summarizes the number of libraries in each group, categorized by lifespan and release speed. \textbf{Our analysis reveals that 59.1\% of abandoned libraries are short-lived, with the majority (68.9\%) exhibiting slow or moderate release speeds.} 20.3\% of abandoned libraries exhibit slow release speed and short-lived, significantly higher than the 8.4\% observed in active libraries. This indicates that short-lived libraries, particularly those with slow release speed, are more prone to abandonment. This trend remains relatively consistent across moderate (1–2 releases per month) and high-speed (\textgreater 2 releases per month) categories, where short-lived abandoned libraries are more prevalent than their active counterparts. 


22.8\% of abandoned libraries are long-lived, suggesting that these libraries may have reached maturity and no longer require frequent updates or new releases. \textbf{Additionally, 43,928 (21.3\%) abandoned libraries exhibit high release speeds. This indicates high release speed alone may not reliably signal the risk of abandonment.}



\subsection{RQ2.2: How does release speed evolve over the lifespan of abandoned libraries compared to active libraries?}


\noindent \textbf{Approach:} The release speed of a library can vary significantly over its lifespan. Thus, in RQ2.2, we perform an in-depth analysis on how a library's release speed evolves across the four quartiles of its lifespan. To ensure meaningful pattern analysis, we include only libraries with at least four releases, providing a sufficiently rich release history for analysis. For each library, we begin by identifying the releases within each of the four quartiles (Q1, Q2, Q3, Q4) based on their release dates. For instance, if the lifespan of a library is 1 year, then releases made in the first three months belong to Q1. Next, we calculate the average lifespan of releases within each quartile and categorize it as ``Fast'', ``Normal'', or ``Slow'' by comparing the quartile-specific lifespan against the overall average lifespan of releases for that library. Specifically, a quartile is labeled as \textit{``Fast''} if its average release lifespan is 20\% less than the library’s overall average.
A quartile is labeled as \textit{``Slow''} if its average release lifespan is 20\% longer than the library’s overall average. Quartiles with intervals between these thresholds are labeled as \textit{``Normal''}. In two cases, quartile-specific average release lifespans cannot be calculated and are labeled as \textit{``nan''}: in either Q2 or Q3, if no releases occur, and in Q4, if only the final release exists, as it coincides with the end of the library's lifespan.



\noindent \textbf{Results:} Tables~\ref{tab:quartile_abandoned_libraries} and~\ref{tab:quartile_active_libraries} present the top-10 sequence patterns observed in abandoned and active libraries, covering 18.2\% of total abandoned libraries and 26.2\% of total active libraries, respectively.



\begin{table}[h]
\centering
\tiny
\caption{Top 10 identified patterns in abandoned libraries.}
\vspace{-0.2cm}
\label{tab:quartile_abandoned_libraries}
\resizebox{\columnwidth}{!}{%
\begin{tabular}{ccc}
\hline
\textbf{\# of libraries} & \textbf{(\%)} & \textbf{Patterns} \\ \hline
6,540 & 5.7 & Fast \textgreater Slow \textgreater nan \textgreater nan       \\
5,428 & 4.7 & Slow \textgreater nan \textgreater nan \textgreater Fast       \\
4,233 & 3.7 & Normal \textgreater nan \textgreater nan \textgreater nan      \\
4,196 & 3.6 & Fast \textgreater Slow \textgreater nan \textgreater Fast      \\
3,651 & 3.2 & Fast \textgreater Slow \textgreater Slow \textgreater nan      \\
2,991 & 2.6 & Fast \textgreater Normal \textgreater Slow \textgreater nan    \\
2,831 & 2.4 & Fast \textgreater Normal \textgreater Slow \textgreater Slow   \\
2,625 & 2.3 & Fast \textgreater Slow \textgreater Slow \textgreater Fast     \\
2,566 & 2.2 & Normal \textgreater Normal \textgreater Normal \textgreater Normal \\
2,487 & 2.2 & Slow \textgreater nan \textgreater Fast \textgreater Fast      \\ \hline
\end{tabular}%
}
\end{table}

\begin{table}[h]
\centering
\tiny
\caption{Top 10 identified patterns in active libraries.}
\label{tab:quartile_active_libraries}

\resizebox{\columnwidth}{!}{%
\begin{tabular}{ccc}
\hline
\textbf{\# of libraries} & \textbf{(\%)} & \textbf{Patterns} \\ \hline
11,047 & 6.7 & Normal \textgreater Normal \textgreater Normal \textgreater Normal \\
8,032  & 4.9 & Fast \textgreater Normal \textgreater Slow \textgreater Slow       \\
4,914  & 2.9 & Fast \textgreater Slow \textgreater Slow \textgreater Slow         \\
4,622  & 2.8 & Fast \textgreater Slow \textgreater nan \textgreater nan           \\
4,351  & 2.6 & Slow \textgreater nan \textgreater nan \textgreater Fast           \\
3,921  & 2.4 & Fast \textgreater Slow \textgreater Slow \textgreater Fast         \\
3,759  & 2.3 & Slow \textgreater Normal \textgreater Normal \textgreater Fast     \\
3,685  & 2.2 & Fast \textgreater Fast \textgreater Slow \textgreater Slow         \\
3,587  & 2.2 & Fast \textgreater Slow \textgreater nan \textgreater Fast          \\
3,509  & 2.1 & Slow \textgreater Slow \textgreater Fast \textgreater Fast         \\ \hline
\end{tabular}%
}
\end{table}
In abandoned libraries, the most common pattern is \textit{``Fast → Slow → nan → nan''}, which exists in 6,540 libraries (5.7\%). This means there are only releases in the first half of the lifespan before the final release, and the release speed is quickly slowed down from the first to the second quartile. \textbf{This aligns with our intuitions, i.e., abandoned libraries already slowed down or have long relative inactive period before abandonment.} Note that this pattern also exists in active libraries, but accounts for only 2.8\% of the cases. These active libraries may still be new, i.e., having their latest release in the final two years of our observation window but may be prone to abandonment soon. The general slowdown patterns can also be observed in other patterns, such as \textit{``Fast → Slow → Slow → nan''} and \textit{``Fast → Normal → Slow → nan''}. Four out of the top-10 patterns represent slowdown, accounting for 7.8\% of the abandoned libraries. The inactive general pattern (containing ``nan'') can also be observed in patterns such as \textit{``Slow → nan → nan → Fast''}, accounting for 7 out of 10 patterns and 14.3\% of the abandoned libraries. This number is much lower in active libraries, with only three patterns containing ``nan''. Interestingly, there are some cases where the libraries recovered from slow release but still get abandoned, i.e., 3.9\% libraries with patterns such as \textit{``Slow → nan → nan → Fast''} and \textit{``Fast → Slow → Slow → Fast''}. This suggests that a temporary increase in release activity is insufficient to prevent abandonment. 

For active libraries, the pattern \textit{``Normal → Normal → Normal → Normal''} is the most common, representing 6.7\% of all active libraries. This suggests that many active libraries maintain a consistent release speed over their lifespan. Although this pattern is also observed in abandoned libraries, it occurs at a significantly lower rate (2.2\%).

\vspace{-0.1cm}
\begin{tcolorbox}[colframe=white, colback=blue!5, left=1mm, right=1mm, sharp corners]
\textbf{RQ2 Summary:} In general, a slowdown in release speed and periods of relative inactivity can signal the potential abandonment of a library. However, some abandoned libraries exhibit fast release speeds toward the end of their lifespan, suggesting that rapid release alone is not always indicative of sustained maintenance.
\end{tcolorbox}


\section{Threats to Validity} \label{threats}

We define abandoned libraries as those without releases during the last two years of our observation window. While this definition may not fully capture true abandonment, i.e., some libraries may become active again in the future, or developers may maintain repositories without releasing new versions, we argue that using releases as the basis for abandonment is reasonable. This is because developers typically rely on package management systems for downloading libraries, and the lack of new releases may pose risks, even if the source code is maintained elsewhere. Additionally, we find that only 6.6\% of the libraries in our dataset experience a release gap of more than two years during their lifespan, suggesting that reactivation after a two-year period of inactivity is highly unlikely. Thus, we believe that the two-year cutoff does not significantly affect our conclusions. However, this definition and cutoff may not generalize to all ecosystems or account for libraries reaching stability or being replaced by alternatives. Furthermore, our study focuses exclusively on the Maven ecosystem, which may limit the applicability of our findings to ecosystems such as npm or PyPI. External factors, such as industry trends or shifts in technology, could also influence release patterns, which our dataset may not fully capture.

\section{Related Work}\label{relwork}

Existing studies have analyzed software development dynamics within ecosystems~\cite{german2013evolution,plakidas2017evolution,decan2019empirical,polese2022adoption}. Among these, the most relevant to our work are those investigating release patterns and project abandonment (survival) at the ecosystem level. Jaime et al.~\cite{jaime2022preliminary} introduced two metrics, i.e., rhythm (time intervals between releases) and speed (average releases per day), to analyze release dynamics in the Maven ecosystem. We also analyze the release speed, but at a more fine-grained level, quartile, and link it with the abandonment of libraries. Ait et al.~\cite{ait2022empirical} analyzed the survival of 1,127 repositories across four ecosystems, i.e., NPM, R, WordPress, and Laravel, over a six-year period. They found that over half of the projects became inactive (abandoned) within four years, with survival rates dropping below 50\% after five years. Miller et al.~\cite{miller2025understanding} analyzed dependency abandonment in the NPM ecosystem. They found that 15\% of widely-used NPM packages were abandoned within six years. Their analysis also highlighted the significant impact of abandonment on dependent projects, many of which fail to respond effectively. Similar to these studies, we find that abandonment is common within the Maven ecosystem. Unlike them, we also investigate library abandonment by examining their release behaviors.

\vspace{-0.1cm}
\section{Discussion and Conclusion}\label{conclusion}
Library abandonment poses significant challenges for software ecosystems, affecting their reliability and maintainability. Our study shows that a slowdown in release speed, particularly in the later stages of a library's lifecycle, often signals impending abandonment. While not all abandoned libraries exhibit this trend, developers are encouraged to monitor dependency release activity, watching for sudden slowdowns or prolonged inactivity, to address potential risks proactively. The high abandonment rate after two years further highlights the need for vigilance and adaptability in managing dependencies. Our findings underscore the importance of release patterns and library lifecycles in understanding abandonment. Future work should aim to identify predictive indicators and develop tools to track library age and release patterns, offering early warnings for at-risk dependencies. The steady high rate of library abandonment in Maven highlights the need for ecosystem-level interventions. Package platforms could introduce automated alerts for libraries showing signs of abandonment, such as inactivity or reduced release frequency, to support maintainers and minimize abandonment risks.


\noindent \textbf{Acknowledgment:} We acknowledge the support of the Natural Sciences and Engineering Research Council of Canada (NSERC), [funding reference number: RGPIN-2019-05071].

\bibliographystyle{IEEEtran}

\bibliography{main}

\begin{thebibliography}{10}
\providecommand{\url}[1]{#1}
\csname url@samestyle\endcsname
\providecommand{\newblock}{\relax}
\providecommand{\bibinfo}[2]{#2}
\providecommand{\BIBentrySTDinterwordspacing}{\spaceskip=0pt\relax}
\providecommand{\BIBentryALTinterwordstretchfactor}{4}
\providecommand{\BIBentryALTinterwordspacing}{\spaceskip=\fontdimen2\font plus
\BIBentryALTinterwordstretchfactor\fontdimen3\font minus \fontdimen4\font\relax}
\providecommand{\BIBforeignlanguage}[2]{{%
\expandafter\ifx\csname l@#1\endcsname\relax
\typeout{** WARNING: IEEEtran.bst: No hyphenation pattern has been}%
\typeout{** loaded for the language `#1'. Using the pattern for}%
\typeout{** the default language instead.}%
\else
\language=\csname l@#1\endcsname
\fi
#2}}
\providecommand{\BIBdecl}{\relax}
\BIBdecl

\bibitem{lin2017developer}
B.~Lin, G.~Robles, and A.~Serebrenik, ``Developer turnover in global, industrial open source projects: Insights from applying survival analysis,'' in \emph{2017 IEEE 12th International Conference on Global Software Engineering (ICGSE)}.\hskip 1em plus 0.5em minus 0.4em\relax IEEE, 2017, pp. 66--75.

\bibitem{forsgren20212020}
N.~Forsgren, B.~Alberts, K.~Backhouse, G.~Baker, G.~Cecarelli, D.~Jedamski, S.~Kelly, and C.~Sullivan, ``2020 state of the octoverse: Securing the world's software,'' \emph{arXiv preprint arXiv:2110.10246}, 2021.

\bibitem{kaur2022exploring}
R.~Kaur and K.~K. Chahal, ``Exploring factors affecting developer abandonment of open source software projects,'' \emph{Journal of Software: Evolution and Process}, vol.~34, no.~9, p. e2484, 2022.

\bibitem{zimmermann2019small}
M.~Zimmermann, C.-A. Staicu, C.~Tenny, and M.~Pradel, ``Small world with high risks: A study of security threats in the npm ecosystem,'' in \emph{28th USENIX Security symposium (USENIX security 19)}, 2019, pp. 995--1010.

\bibitem{miller2023we}
C.~Miller, C.~K{\"a}stner, and B.~Vasilescu, ``“we feel like we’re winging it:” a study on navigating open-source dependency abandonment,'' in \emph{Proceedings of the 31st ACM Joint European Software Engineering Conference and Symposium on the Foundations of Software Engineering}, 2023, pp. 1281--1293.

\bibitem{valiev2018ecosystem}
M.~Valiev, B.~Vasilescu, and J.~Herbsleb, ``Ecosystem-level determinants of sustained activity in open-source projects: A case study of the pypi ecosystem,'' in \emph{Proceedings of the 2018 26th ACM Joint Meeting on European Software Engineering Conference and Symposium on the Foundations of Software Engineering}, 2018, pp. 644--655.

\bibitem{avelino2019abandonment}
G.~Avelino, E.~Constantinou, M.~T. Valente, and A.~Serebrenik, ``On the abandonment and survival of open source projects: An empirical investigation,'' in \emph{2019 ACM/IEEE International Symposium on Empirical Software Engineering and Measurement (ESEM)}.\hskip 1em plus 0.5em minus 0.4em\relax IEEE, 2019, pp. 1--12.

\bibitem{coelho2020github}
J.~Coelho, M.~T. Valente, L.~Milen, and L.~L. Silva, ``Is this github project maintained? measuring the level of maintenance activity of open-source projects,'' \emph{Information and Software Technology}, vol. 122, p. 106274, 2020.

\bibitem{cogo2021deprecation}
F.~R. Cogo, G.~A. Oliva, and A.~E. Hassan, ``Deprecation of packages and releases in software ecosystems: A case study on npm,'' \emph{IEEE Transactions on Software Engineering}, vol.~48, no.~7, pp. 2208--2223, 2021.

\bibitem{ait2022empirical}
A.~Ait, J.~L.~C. Izquierdo, and J.~Cabot, ``An empirical study on the survival rate of github projects,'' in \emph{Proceedings of the 19th International Conference on Mining Software Repositories}, 2022, pp. 365--375.

\bibitem{miller2025understanding}
C.~Miller, M.~Jahanshahi, A.~Mockus, B.~Vasilescu, and C.~K{\"a}stner, ``Understanding the response to open-source dependency abandonment in the npm ecosystem,'' in \emph{Int’l Conf. Software Engineering (ICSE), IEEE/ACM}, 2025.

\bibitem{shen2025understanding}
Y.~Shen, X.~Gao, H.~Sun, and Y.~Guo, ``Understanding vulnerabilities in software supply chains,'' \emph{Empirical Software Engineering}, vol.~30, no.~1, pp. 1--38, 2025.

\bibitem{msrdata}
D.~Jaime, J.~El~Haddad, and P.~Poizat, ``Navigating and exploring software dependency graphs using goblin,'' in \emph{Proceedings of the International Conference on Mining Software Repositories (MSR 2025)}, 2025.

\bibitem{10.1145/3643991.3644879}
\BIBentryALTinterwordspacing
D.~Jaime, J.~E. Haddad, and P.~Poizat, ``Goblin: A framework for enriching and querying the maven central dependency graph,'' in \emph{Proceedings of the 21st International Conference on Mining Software Repositories}, ser. MSR '24.\hskip 1em plus 0.5em minus 0.4em\relax New York, NY, USA: Association for Computing Machinery, 2024, p. 37–41. [Online]. Available: \url{https://doi.org/10.1145/3643991.3644879}
\BIBentrySTDinterwordspacing

\bibitem{jaime2022preliminary}
D.~Jaime, J.~El~Haddad, and P.~Poizat, ``A preliminary study of rhythm and speed in the maven ecosystem,'' in \emph{21st Belgium-Netherlands Software Evolution Workshop}, 2022.

\bibitem{german2013evolution}
D.~M. German, B.~Adams, and A.~E. Hassan, ``The evolution of the r software ecosystem,'' in \emph{2013 17th European Conference on Software Maintenance and Reengineering}.\hskip 1em plus 0.5em minus 0.4em\relax IEEE, 2013, pp. 243--252.

\bibitem{plakidas2017evolution}
K.~Plakidas, D.~Schall, and U.~Zdun, ``Evolution of the r software ecosystem: Metrics, relationships, and their impact on qualities,'' \emph{Journal of Systems and Software}, vol. 132, pp. 119--146, 2017.

\bibitem{decan2019empirical}
A.~Decan, T.~Mens, and P.~Grosjean, ``An empirical comparison of dependency network evolution in seven software packaging ecosystems,'' \emph{Empirical Software Engineering}, vol.~24, no.~1, pp. 381--416, 2019.

\bibitem{polese2022adoption}
A.~Polese, S.~Hassan, and Y.~Tian, ``Adoption of third-party libraries in mobile apps: a case study on open-source android applications,'' in \emph{Proceedings of the 9th IEEE/ACM International Conference on Mobile Software Engineering and Systems}, 2022, pp. 125--135.

\end{thebibliography}
\end{document}